\documentstyle[11pt,newpasp,twoside]{article}
\def\edcomment#1{\iffalse\marginpar{\raggedright\sl#1\/}\else\relax\fi}
\reversemarginpar

\begin{document}
\title{The dusty SF history of distant galaxies and modelling tools}
 \author{Gian Luigi Granato}
\affil{Osservatorio Astronomico di Padova - Vicolo Osservatorio 5 
- I35122 Padova - Italy}

\begin{abstract}

I review recent advances in the determination of the cosmic history of star
formation, and its relevance in our understanding of the formation of
structures. I emphasize the importance of dust reprocessing in the high--z
universe, as demonstrated in particular by IR and sub-mm data. This demand a
panchromatic approach to observations and suitable modelling tools. I summarize
the basic requirements for these models and to what extent they are satisfied
by models published so far.

\end{abstract}

\section{Introduction}

In the last few years, a remarkable number of studies of the high-z universe
have been devoted to the determination of the cosmic history of star formation
$SFR(z)$. The main motivation for these efforts is that baryons are the only
observational tracers of the evolution of large scale structures, which are
driven by the gravitational collapse of dark matter (DM). The function $SFR(z)$
in principle constrains cosmogonic theories for the build up of structures,
with the obvious, but sometimes underestimated, caveat that baryons assembly is
strongly affected also by a much more complex physics than dark matter.

In the local universe surveyed by IRAS $\sim 30\%$ of the energy is dust
reprocessed: if the same were true at high--z, optical--UV observations would
suffice to determine $SFR(z)$ with a small uncertainty. But the IRAS
observations demonstrated as well that in local galaxies dust reprocessing is an
increasing function of the star formation activity, and can't be reliably
determined by UV and optical data alone. This is vividly illustrated for
instance by figure 2 of Sanders and Mirabel (1996), which shows that
infrared-selected galaxies range over 3 order of magnitude in $L_{ir}$, while
the optical luminosity change only by a factor of 3-4, with minor differences
in  the shape of the optical-UV SED.

It is therefore quite natural to suspect that in the more active young universe
an higher fraction of star luminosity was reprocessed by dust.




\section{The dusty SF history of galaxies}

Indeed, several pieces of evidence have shown that {\it most of the SF in
the high--z universe is dust obscured or dimmed to a substantial  degree:} (1)
The discovery of a cosmic far-IR/sub-mm background by the COBE satellite (Puget
et al 1996, Fixsen et al 1998, Hauser et al 1998), whose energy density, 
which is at least a factor 2-3 larger than the optical-UV one,
indicates that a large fraction of the energy radiated by stars over the
history of the universe has been reprocessed by dust.  (2) The discovery that
the population of star forming galaxies at $z\sim 2-4$ that have been detected
through their strong Lyman-break features are substantially extincted in the
rest-frame UV (Pettini et al 1998, Steidel et al 1999). (3) The discovery of a
population of sub-mm sources at high redshift ($z> 1$) using SCUBA, whose
luminosities, if they are powered by star formation in dust-enshrouded
galaxies, imply very large star formation rates ($\sim 10^2
M_{\odot}\mbox{yr}^{-1}$), and a total star formation density comparable to
what is inferred from the UV luminosities of the Lyman-break galaxies (Smail et
al 1997, Hughes et al 1998, Lilly et al 1999). (4) The ISO detection of a
population of strong IR sources; 15 $\mu$m ISOCAM (Oliver et al 1997) and 175
$\mu$m ISOPHOT surveys (Kawara et al 1998, Puget et al 1999) show a
population of actively star forming galaxies at $0.4 < z < 1.3$, which
boosts the cosmic star formation density by a factor $\sim 3$ with respect to
that estimated in the optical from the CFRS. For (1) and (3),
there is the caveat that the contribution from dust-enshrouded AGNs to the
sub-mm counts and background is currently uncertain, but probably the AGNs do
not dominate (e.g.\ Granato, Danese \& Franceschini 1997).

A summary of the present status of the determination of $SFR(z)$ is given for
instance by figure 17 of Genzel and Cesarsky (2000). The main point is that while a
few years ago it was claimed, based on optical observation, that this function
has a peak at $z\simeq 1$ and declines at higher redshift, it is now clear
instead that $SFR(z)$ steeply increases between $z=0$ and $z=1$, but than stays
flat to at least $z\simeq 4$.

\section{Implications}

As already mentioned, the determination of $SFR(z)$ aims to constrain
cosmogonic scenarios. In the past decades, two extreme opposite possibilities
have been studied:

(1) {\it monolithic models} (e.g.\ Eggen, Lynden-Bell \& Sandage 1962, Larson
1975, etc), characterized by a timetable for SF dependent on morphological
type, and by an evolution of galaxies as individual units. In particular,
ellipticals had a huge burst of SF at high z, followed by a monolithic/passive
evolution.

(2) By converse, according to {\it hierarchical clustering models} (e.g.\ White \&
Rees 1978, White \& Frenk 1991), massive objects formed at late times by the
merging of smaller subunits, and therefore the assembly of massive ellipticals
occurred at $z< 1$;

The early claims of a $SFR(z)$ peaking around $z\sim 1$ were warmly welcome as
a success of the prediction of hierarchical clustering models, but as already
remarked the observational situation has now changed.

Without going into the details of the evidences in favor or against the two
scenarios above, it seems now that clear they are converging to some
intermediate point. Indeed monolithic models have been originally developed
focusing on astrophysical arguments concerning luminous matter. By converse,
the merging picture has been mostly driven by the predicted evolution of dark
matter halos through hierarchical clustering. For instance, on one hand, a more
prolonged SF for field ellipticals has been proposed (e.g.\ Franceschini et
al.\ 1998) and a role of merging below $z\simeq 1$ is apparent from optical
surveys (Schade et al.\ 1999 and references therein). On the other hand present
semi-analytical models, while reproducing several observations, are seriously
at odd with SCUBA counts (Silva 1999), and do not reproduce the observed trends
of $\alpha$-enhancement in ellipticals (Thomas \& Kauffmann, 2000).

The emerging picture (say the {\it "middleman" scenario}, e.g.\ Schade et al
1999) is that massive ellipticals assembly most of their stellar mass early
(e.g. $z> 3$) but some fraction of stars form later ($z<2$). To obtain this,
the simplified prescriptions adopted by semi-analytical models to describe the
complex behavior of baryons in DM halos need some revision.

\section{Modelling tools}

In any case, the recent discoveries demonstrate that in order to understand the
history of star formation in the universe from observational data, a unified
picture, covering all wavelengths from the far-UV to the sub-mm, is
indispensable. The UV and the far-IR are especially important, since young
stellar populations emit most of their radiation in the rest-frame UV, but a
significant fraction of this is dust reprocessed into the rest-frame far-IR.
Therefore (i) {panchromatic observations}, with proper emphasis on the mid-IR
to the sub-mm, are required and (ii) these observations should be interpreted
using {spectral synthesis tools taking into account dust reprocessing} in a
decorous way.

The spectral energy distributions (SEDs) of model galaxies are computed with
the spectral synthesis technique. If dust were negligible in the system, as it
is usually assumed to be the case in local elliptical galaxies, the only
required ingredients would be {(1)} the distribution of stars in age and
metallicity, and {(2)} a library of Simple Stellar Population spectra (SSPs),
i.e.\ the integrated spectrum of a single generation of stars of given initial
mass function (IMF), age and metallicity.
 
Then the galaxy SED at a given age $t_G$ would be given by a simple integral
over its past history. For a monolithic evolution model, were there is a one to
one relationship $Z(t)$ between the galactic age and its metallicity, we 
have:

\begin{equation}
L_\lambda(t_G)=\int_0^{t_G} SSP_\lambda(t_G-t,Z(t))\times SFR(t)\:dt 
\end{equation}

\noindent were $SFR(t)$ is the star formation rate.

In merging models there is no unique age-metallicity relation $Z(t)$, since
each galaxy results from the 'combination' of several progenitor galaxies which
have merged to produce that galaxy. The progenitor galaxies each had their own
star formation and chemical history. In this case a birthrate function
$\Psi(t,Z)$ is introduced, where $\Psi(t,Z)\,dt\, dZ$ gives the mass of stars
that were formed in the time interval $(t,t+dt)$ with metallicities in the
range $(Z,Z+dZ)$. The composite $\Psi(t,Z)$ in general has a broad distribution
of metallicity at each age, and the above integral has to be replaced by a
slightly more complicated computation:

\begin{equation}
L_\lambda(t_G)=\int dZ \int_0^{t_G} SSP_\lambda(t_G-t,Z)\times \Psi(t,Z)\:dt 
\end{equation}

But the true complexity arises whenever dust reprocessing must be taken into
account, which is the rule for star forming systems. Dust absorbs and scatters
photons very effectively below $\sim 1 \mu$m, and thermally reradiates the
absorbed energy above a few $\mu$m. This yields large modifications in the
SED. These effects require radiative transfer computation of starlight
through dust. This is by itself a major complication, since in any geometry
with a minimum of realism the radiative transfer can be done only by means of
numerical techniques. But the worst problem is that the results are a strong
function of the optical properties of dust and of the geometrical arrangement
of dust and stars, introducing several parameters and uncertainties
in the models.

Concerning the dust optical properties, they are relatively well understood in
our own galaxy, but they are known to be a function of the environment. At
least three different sites of interaction between stellar photons and dust
grains need separate consideration: AGB envelopes, molecular clouds (MCs),
diffuse ISM. Moreover variations of dust properties even within each of these
environments are known to exist. Further complications come from the presence
of small grains, which are not in thermal equilibrium with the radiation field
and whose emission has to be computed by means of statistical techniques,
and/or by modelling the carriers of unidentified IR bands, likely Policiclyc
Aromatic Hydrocarbons.

As for the geometry of stars and dust, it is certainly much more complex than
that adopted in most models. At very least two phases in the ISM should be
considered: besides a relatively smooth diffuse ISM, responsible in the Galaxy
of the cirrus emission discovered by IRAS, a clumpy component, the MCs, is
present. Moreover, since the birth and first evolutionary stages of stellar
lives occur in MCs, the 'primary source' of radiation is clumpy as well, since
the younger the stars the more closely associated to MCs they are.

Several papers in the last few years tried to address some of these technical
or astrophysical difficulties, which are on the other hand the minimal
requirements for a meaningful modelling of UV to sub-mm SED of star forming
galaxies. A non exhaustive briefly commented list of most recent models
follows:

\begin{itemize} 

\item Witt, Thronson \& Capuano (1992), Gordon, Calzetti \& Witt (1997), Fioc
\& Rocca Volmerange (1997, PEGASE), Ferrara et al.\ (1999), Bianchi et al.\
(2000) considered only the extinction effects of dust. These models, with the
only exception of PEGASE, use Monte Carlo methods, allowing a treatment of
scattering with an accuracy difficult to reach with other methods, which
however tend to be much faster. Monte Carlo computations are therefore
particularly suited for detailed comparison with optical images of single
spiral galaxies. However in all but Bianchi et al.\ (2000) paper the assumed
geometry is more or less oversimplified with respect to real galaxies, lacking
in particular the consideration of molecular clouds and their association with
young stars.

\item Other authors (e.g.\ Mazzei, De Zotti \& Xu 1994; Devriendt, Guiderdoni
\& Sadat 1999: STARDUST) included also a computation of dust emission, but with
substantial simplifications. In the latter paper, dust absorption is modelled
assuming a 1D slab geometry, and the dust temperature distribution is not
predicted. Instead, the dust emission spectrum is modelled as the sum of
several components, whose temperatures and relative strengths are chosen so as
to reproduce the observed correlations of IR colors with IR luminosity found
by IRAS in the local Universe. While this kind of approach is undoubtedly very
simple and has relatively few parameters, it is not directly linked to the
physics of dust emission. Therefore it lacks of predictive power for systems
very different from the local galaxies against which it has been calibrated. In
particular it may be limitative for high redshift applications.

\item Efstathiou, Rowan--Robinson \& Siebenmorgen (2000) and  Bianchi et al.\
(1999) also included dust emission, but still with some restrictions. The first
paper introduces a simple physical treatment of the time evolution of GMC,
but, being thought for applications to starburst galaxies, it neglects the
effects of diffuse dust. These are important at least for normal spirals but to
some level even for weak starbursts. The latter paper represents probably the
state of the art of Monte Carlo models in this field. The only limitation,
besides the intrinsic slowness of the method, is that small grains and PAH are
not yet included, preventing the comparison with MIR data.

\end{itemize}

The only general purpose model in which all the geometric and optical
properties issues listed above have been addressed is GRASIL (Silva, Granato,
Bressan \& Danese 1998, executables are available for download at {\it
http://grana.pd.astro.it/}). This stellar population + dust model includes a
realistic 3D geometry, with a disk and bulge, two phase dust in clouds and in
the diffuse ISM, star formation in the clouds, radiative transfer of starlight
through the dust distribution, a realistic dust grain model including PAHs and
quantum heating of small grains, and a direct prediction of the dust
temperature distribution at each point in the galaxy based on a calculation of
dust heating and cooling. The adopted algorithm allows its use in applications
requiring predicted SEDs from the UV to the sub-mm of thousands of galaxies, in
reasonable computing time (e.g.\ Granato et al.\ 2000). The comparison with
results from more precise Monte Carlo computations is on the other hand very
positive (for details see {\it http://grana.pd.astro.it/}). GRASIL is the first
published model taking into account a progressive escape of young stars from
the thickest phase of the ISM, namely the parent molecular clouds. This feature
has proven to be fundamental to understand UV properties of galaxies, in
particular the shallow and featureless attenuation law of starburst galaxies
(see Granato et al.\ 2000 for details).


\begin{references}

\reference Bianchi, S., Ferrara, A., Davies, J. I. and Alton, P. B. 2000, \mnras, 311, 
601 
\reference Bianchi, S., Davies, J. I. and Alton, P. B. 2000, accepted by \aap, 
(astro-ph/0005103)
\reference Devriendt, J. E. G., Guiderdoni, B. \& Sadat, R. 1999,
\aap, 350, 381 
\reference Efstathiou, A., Rowan-Robinson, M. \& 
Siebenmorgen, R. 2000, \mnras, 313, 734 
\reference Eggen, O. J., Lynden-Bell, D. and Sandage, A. R. 1962, \apj, 136, 748 
\reference Ferrara, A., Bianchi, S., Cimatti, A. \&
Giovanardi, C. 1999, \apjs, 123, 437 
\reference Fioc, M. \& Rocca-Volmerange, B. 1997, \aap, 326, 950 
\reference Fixsen, D.J., Dwek, E., Mather, J.C., Bennett, C.L., Shafer, R.A.,
1998, \apj, 508, 123
\reference Franceschini, A., Silva, L., Fasano, G., Granato, L., Bressan, A.,
Arnouts, S., Danese, L. 1998, \apj, 506, 600 
\reference Genzel, R. \& Cesarsky, C. 2000, \araa, in press (astro-ph/0002184)
\reference Gordon, K. D., Calzetti, D. \& Witt, A. N. 1997, \apj, 487, 625 
\reference Granato, G.L., Danese, L. \& Franceschini, A., \apj, 486, 147
\reference Granato, G.L. Lacey, C.G., Silva, L., Bressan, A., Baugh, C.M.,
Cole, S. \& Frenk, C. 2000, to appear on 20 October 2000 issue of ApJ
(astro-ph/0001308)
\reference Hauser, M.G., et al. 1998, \apj, 508, 25
\reference Hughes, D.H., et al. 1998, nature, 394, 241
\reference Kawara, K., et al. 1998, \aap, 336, L9 
\reference Larson, R. B. 1975, \mnras, 173, 671 
\reference Lilly, S.J., LeFevre, O., Hammer, F., Crampton, D., \apj, 460, L1
\reference Mazzei, P., de Zotti, G. \& Xu, C. 1994, \apj, 422, 81 
\reference Oliver, S. J., et al. 1997, \mnras, 289, 471 
\reference Pettini, M., Kellogg, M., Steidel, C., Dickinson, M., Adelberger,
K.L., Giavalisco, M., 1998, \apj, 508, 539
\reference Puget, J.L, et al. 1996, \aap, 308, L5
\reference Puget, J. L., et al. 1999, \aap, 345, 29 
\reference Sanders, D.B. and Mirabel I.F. 1996, \araa, 34, 749
\reference Schade, D., et al.\ 1999, \apj, 525, 31 
\reference
Silva, L., Granato, G. L., Bressan, A. \& Danese, L. 1998, \apj, 509, 103 
\reference Silva, L., 1999, SISSA PhD thesis, (http://grana.pd.astro.it)
\reference Smail, I., Ivison, R.J., Blain, A.W., 1997, \apj, 490, L5
\reference Steidel, C.C., Adelberger, K.L., Giavalisco, M., Dickinson, M.,
Pettini, M., 1999, \apj, 519, 1 
\reference Thomas, D., \& Kauffmann, G. 1999, to appear in the 
PASP Conference Proccedings
"Spectrophotometric dating of stars and galaxies",
Annapolis, Maryland (USA), eds. I. Hubeny, S. Heap, and R. Cornett
(astro-ph/9906216)
\reference White, S. D. M. \&  Rees, M. J. 1978, \mnras, 183, 341 
\reference White, S. D. M. \& Frenk, C. S. 1991, \apj, 379, 52 
\reference Witt, A. N., Thronson, H. A. and Capuano, J. M. 1992, \apj, 393, 611 


\end{references}
\end{document}